\documentclass[twocolumn,preprintnumbers,amsmath,amssymb]{revtex4}
\usepackage{graphicx}
\usepackage{dcolumn}
\usepackage{bm}% bold math

\begin{document}
\title{Population size bias in Diffusion Monte Carlo}
\author{Massimo Boninsegni\footnote{m.boninsegni@ualberta.ca}$^{1,2}$ and Saverio Moroni$^2$}
\affiliation{$^1$\ Department of Physics, University of Alberta, Edmonton, Alberta, Canada, T6G 2G7}
\affiliation{$^2$\ SISSA Scuola Internazionale Superiore di Studi Avanzati and DEMOCRITOS National 
Simulation Center,
Istituto Officina dei Materiali del CNR Via Bonomea 265, I-34136, Trieste, Italy}
\date\today

\begin{abstract}
The size of  the population of random walkers required to obtain converged  estimates in DMC increases dramatically with  system size. We illustrate this by comparing ground state energies of small clusters of parahydrogen (up to 48 molecules) computed by 
Diffusion Monte Carlo (DMC) and Path Integral Ground State (PIGS) techniques.  We contend that the bias associated to a finite population of walkers is the most likely cause of 
quantitative numerical discrepancies between PIGS and DMC energy estimates reported in the literature, for this few-body Bose system. We 
discuss the viability of DMC as a general-purpose ground state technique, and argue that PIGS, and even finite temperature methods, enjoy more favorable scaling, and are therefore  a superior option for systems of large size.
\end{abstract}

\maketitle

\section{INTRODUCTION}
\label{intro}
Quantum Monte Carlo (QMC)  methods are widely utilized to compute accurate thermodynamics of  quantum 
few-body systems. The best known, and arguably most popular  such method, is the Diffusion Monte Carlo (DMC), 
which has been extensively adopted over the past three decades, especially in the context of electronic 
structure calculations for atoms and molecules \cite{Lester}, but also in studies of light nuclei 
\cite{Pieper}, as well as of small Bose clusters such as ($^4$He)$_N$ \cite{schmidt92} or (H$_2$)$_N$ 
\cite{whaley}.
\hfil\break\indent
On the other hand, the Path Integral Ground State (PIGS) \cite{Ceperley1,Magro,Roy} and related methods 
\cite{baroni}, have only relatively recently emerged as an interesting alternative to DMC. The most 
obvious  advantage of PIGS over DMC is the straightforward, unbiased computation of ground state expectation values 
of quantities other than the energy, including off-diagonal correlations such as the one-body density 
matrix \cite{saverio,holzmann,carleo}, not accessible within DMC. 
\hfil\break\indent
There is, however, another significant difference between the two methods, one that may have so far been overlooked and/or understated in the literature, namely that the results obtained by DMC are intrinsically biased by a necessarily finite population of random walkers. PIGS, on the other hand, is affected by no such limitation; 
we argue in this paper that this  yields PIGS an edge over DMC, as systems of
increasing number of particles are investigated. Specifically, we 
show quantitatively, using a simple test system, that the bias arising from a fixed finite population is a rapidly increasing function of the number $N$ of 
particles in the system (possibly leading to an exponential scaling of the 
computational cost); furthermore, for a given $N$ and for the typical numbers 
of random walkers 
commonly utilized, the bias can be  both suprisingly large in magnitude, as well as difficult to control or remove, as the extrapolation of  results obtained for different population sizes is not only very time-consuming, it can be  quite problematic as well.
\hfil\break\indent
We illustrate the above conclusions by carrying out a systematic  comparison 
of ground state energy estimates yielded by DMC and PIGS, for a small cluster of parahydrogen (H$_2$) 
molecules, including between $N$=13 and $N$=48 molecules.
%This behavior is observed  irrespective on the trial wave function utilized. 
We deem 
this a cogent test case, as a finite Bose cluster could be regarded as the  paradigm 
physical system for which DMC ought to be applicable straightforwardly, almost as a ``black box".
\\ \indent
The remainder of this manuscript is organized as follows: in the next section, we briefly review the 
basic differences between DMC and PIGS. Because both techniques are extensively discussed
in the literature, we refer the reader to the appropriate references for a more in-depth illustration (see, 
for instance, Refs. \onlinecite{Magro,umrigar}).
We then outline the model utilized in this work as a test case to perform calculations, and devote the bulk 
of this paper to a thorough presentation of the
numerical results. We then discuss whether the bias due to a finite walker population may be the (main) cause of
outstanding discrepancies between energy estimates for parahydrogen clusters reported in the literature,
and offer our view on the importance of the population size bias on the scalability of DMC. On this point,
we note that the hypothesis of an overall exponential scaling with $N$ of the computational resources needed for DMC,
has already been put forward by others \cite{nemec}.
\section{Methods}
PIGS and DMC have the same theoretical basis; in both, the exact ground state of a quantum system is 
projected out of an initial  trial state, by simulating on a computer its evolution in imaginary time. 
Consider for definiteness 
a system of $N$ identical particles of mass $m$;  we assume for simplicity that the system obeys 
{\it Bose} statistics \cite{notesign}. \\ \indent 
The quantum-mechanical Hamiltonian $\hat {\cal H}$ 
of the system is 
\begin{equation}\label{ham}
\hat {\cal H}=\hat {\cal H}_\circ + \hat V = -\lambda \sum_{i=1}^N\nabla^2_i + V(R) \end{equation}
where $\lambda={\hbar^2}/{2m}$, $R\equiv {\bf r}_1{\bf r}_2...{\bf r}_N$, are the positions of the 
$N$ particles, and $V(R)$ is the total potential energy of the system associated with the many-particle 
configuration $R$ (this is typically the sum of pairwise interactions, but can be more general). 
The exact ground state wave function $\Phi_\circ(R)$   can be formally  obtained from an initial 
trial wave function $\Psi_T(R)$ as
\begin{eqnarray}\label{proj}
\Phi_\circ(R) \propto {\rm lim}_{\beta\to\infty} \int dR^\prime\ G(R,R^\prime,\beta) \ \Psi_T(R^\prime)
\end{eqnarray}
where \begin{equation}\label{prop}
G(R,R^\prime,\beta) = \langle R | {\rm exp}[-\beta\hat{\cal H}]|R^\prime\rangle
\end{equation}
is commonly referred to as the imaginary-time propagator. While Eq. (\ref{proj}) is formally exact, for a 
nontrivial many-body problem one does not normally have access to $G(R,R^\prime,\beta) $. 
However,  using one of several available schemes, it is possible to obtain approximations for $G$, whose 
accuracy increases as $\beta\to 0$;  if $G_\circ(R,R^\prime,\beta)$ is one such approximation, one can 
take advantage of the identity ${\rm exp}[-\beta\hat H]\equiv ({\rm exp}[-\tau\hat H] )^M$, 
with $\beta=M\tau$, and obtain  $G(R,R^\prime,\beta)$ as 
\begin{eqnarray}\label{path}
G(R,R^\prime,\beta) \approx \int \ \prod_{i=0}^{M-1}\ dR_i\ \biggl\{ G_\circ(R_{i+1},R_i,\tau)
\biggr \} 
\end{eqnarray}
where $R\equiv R_0, \ R_M\equiv R^\prime$.
Eq. (\ref{path}) is exact in the limit $M\to\infty$ (i.e., $\tau\to 0$), which can be achieved in 
practice by extrapolating numerical results obtained with different values of $M$. 

The difference between PIGS and DMC lies in how the above procedure is implemented numerically. 
In PIGS, one generates sequentially, on a computer, a large set $\{X^p\}$, $p=1,2,...,P$, of many-particle 
paths $X\equiv R_0 R_1\ ...\
R_{2M}$ through configuration space. Each $R_j \equiv {\bf r}_{j1}{\bf
r}_{j2} \ ... \ {\bf r}_{jN}$ is a point in 3$N$-dimensional space,
representing positions of the $N$ particles in the
system. These paths are statistically sampled, using the Metropolis algorithm, from a probability density
\begin{eqnarray} 
{\cal P}(X)\propto \Psi_T(R_0)\Psi_T(R_{2M}) \
 \biggl \{ \prod_{i=0}^{2M-1} G_\circ(R_{i+1},R_i,\tau)\biggr \}\label{pippo}
\end{eqnarray}

It is a simple matter to show \cite{Ceperley1,Magro} that in the limits $\tau
\to 0$, $M\tau \to \infty$, $R_{M}$ is sampled from a probability density
proportional to the square of the exact ground state wave function $\Phi_\circ(R)$, {\it irrespective of the choice of} 
$\Psi_T$ \cite{notey}. One can therefore
use the set $\{R_M^p\}$ of ``midpoint" configurations $R_{M}$ of the statistically sampled paths, to compute ground 
state expectation values of thermodynamic quantities $F(R)$ that are diagonal in the position
representation, simply as statistical averages, i.e.
\begin{equation}\label{pix}
\langle\Phi_\circ | \hat F(R)|\Phi_\circ\rangle \approx \frac{1}{P}\ \sum_{p=1}^P\
F(R_M^p),
\end{equation} 
an approximate equality, asymptotically exact in the $P \to
\infty$ limit. The ground state expectation value of the energy can be
obtained in several ways; it is particularly convenient to use the ``mixed estimate"
\begin{equation}\label{mix}
\langle\Phi_\circ|\hat H|\Phi_\circ\rangle 
=\langle\Phi_\circ|\hat H|\Psi_T\rangle 
\approx \sum_{p=1}^P\ \frac {\hat H\Psi_T(R_1^p)}{\Psi_T(R_1^p)}
\end{equation}
which provides an unbiased result for the Hamiltonian operator $\hat H$.
\hfil\break
Obviously, the total projection time $\beta\equiv M\tau$ remains finite. It is straightforward to prove that the energy estimate $E(\beta)$, corresponding to  a finite value of $\beta$ is a strict upper bound on the exact ground state energy $E_\circ$, which is approached monotonically in the $\beta\to\infty$ limit as  
\begin{equation}E(\beta)-E_\circ\sim c\exp(-\beta\ \Delta E),\end{equation} 
where $\Delta E$ is the energy gap between the ground state and the first excited state.
\hfil\break\indent
By contrast, DMC implements the imaginary time evolution of the initial, trial state $\Psi_T$ 
by introducing an importance-sampling transformation of Eq. (\ref{proj}),
\begin{eqnarray}\label{is_proj}\nonumber 
&\Phi_\circ(R)\Psi_G(R) \propto   \\ 
&{\rm lim}_{\beta\to\infty} \int dR^\prime\ {\tilde G}(R,R^\prime,\beta) \ \Psi_T(R^\prime)\Psi_G(R^\prime),  
\end{eqnarray}
where $\Psi_G$ is a positive-definite guidance function
and $ {\tilde G}(R,R^\prime,\beta)=\Psi_G(R)G(R,R^\prime,\beta)/\Psi_G(R^\prime)$.
Hereafter, as almost invariably done for Bose systems, we take $\Psi_T=\Psi_G$. 
Eq. (\ref{is_proj}) is simulated by a guided, diffusive random walk 
through configuration space of a population of $N_W$ (ideally uncorrelated) walkers. Each walker performs successive
transitions from its present configuration $R_p$ to a new one $R_n$, sampled from a diffusive  probabilistic kernel contained in ${\tilde G}_\circ(R_n,R_p,\tau)$, with 
the addition of a drifting term which depends on $\Psi_T$.
The aim of such a drifting term 
is allowing for importance sampling of the configurations, normally expected to reduce considerably the variance 
of the estimates. There is no importance sampling in PIGS, on the other hand \cite{Roy,reatto}.
\hfil\break\indent
A crucial feature of DMC 
is the fact that walkers, along the random walk, accumulate weights proportional to 
${\rm exp}[-\int d\tau E_L(\Psi_T(R))\tau)]$, where 
$E_L(\Psi_T(R))\equiv\hat{\cal H}\Psi_T(R)/\Psi_T(R)$ is the local energy given by the trial wave function 
at the configuration $R$, visited at imaginary time $\tau$ by a given walker. Typically, weights fluctuate considerably, 
both along the random walks, as well as within the population at any given time. Therefore, it proves convenient to 
reconfigure the population, every now and then during the calculation, so that walkers whose weights have become 
negligibly small are discarded, and copies are made of walkers whose weights are larger. This reconfiguration, known as 
{\it branching}, is done in such a way that fluctuations in the weights of individual walkers remain limited.
%SM scritto un pochino piu' esplicitamente da dove viene il bias
%   (cioe' non il branching ma il controllo della popolazione).
In addition, control must be exerted in order to limit fluctuations in the 
population size (or total weight). 
Here too, it is possible to show that in the limit of long 
projecion time the population of walkers will sample a distribution of configurations proportional to  $\Phi_\circ(R)\Psi_T(R)$, which can then be used to evaluate the exact ground state energy.
\hfil\break\indent
The main advantage of this computational strategy, at least in principle, is that the projection time can be made very 
large with little computational effort. On the other hand,  a bias is introduced in the procedure, as one must 
necessarily work with a finite population of walkers. In order for the algorithm to be exact,  extrapolation to infinite population size must be carried out (see, for instance, Ref. \onlinecite{umrigar} for details).
\hfil\break\indent
There has been surprisingly  little work aimed at establishing the magnitude of the finite population size bias  on the computed expectation 
values, but some calculations have shown that  it can be significant, particularly when trying to estimate expectation 
values of operators that do not commute with the Hamiltonian \cite{runge92,calandra,ceperleylast}. On general grounds, one can expect the bias to depend on the accuracy of the guiding wave function $\Psi_T$; if, hypothetically, the {\it exact} ground state wave function were known, then a single walker would suffice, as branching would disappear and the DMC calculation would reduce to a  variational one, as expected. On the other hand, the less accurate $\Psi_T$, the more significant the fluctuations of the local energy associated to individual walkers, and with those the more important the effect of  branching, from which the need for a larger population size ensues.
Within PIGS there is no such bias, as there is no population and no branching.

\section{Model and calculations}\label{mod}

Our physical system of interest, for which we present all the numerical results discussed here, 
is a self-bound cluster of   $N$ parahydrogen molecules, regarded as point particles,
moving in three dimensions. This system has been the subject of much theoretical investigation over the past few years, as it is believed to display an interesting interplay of classical and quantum-mechanical physical effects \cite{holland}.  Clusters of parahydrogen of less than 20 molecules are liquidlike and superfluid at low $T$;  if the number of molecules is between 20 and 40, clusters can ``quantum melt" at low temperature  i.e., go from a solidlike arrangements, with molecules sitting at preferred sites, to a superfluid one, in which they are essentially delocalized throughout the cluster \cite{mezzacapo, mezzacapo1}.  Moreover, some specific ``supersolid" clusters, superfluid and solid behaviours appear to coexist in the $T\to 0$ limit \cite{supersolid}.
\hfil\break\indent
An interesting issue is whether there exist clusters of specific sizes (also referred to as ``magic numbers") that enjoy enhanced stability  over others. This has been investigated in a number of works by computation of the total ground state energy $E(N)$, as a function of cluster size $N$, and by looking for isolated peaks of the chemical 
potential $\mu(N)$, defined as
\begin{equation}\label{muc}
\mu(N) = E(N-1)-E(N)
\end{equation}
Clearly, the precise identification of magic clusters requires a sufficiently accurate determination of $E(N)$, which is an extensive quantity.
At present, there exist outstanding discrepancies between different ground state results obtained by DMC \cite{guardiola, sola}, PIGS \cite{cuervo}, as well as by extrapolating to $T=0$ results at finite temperature \cite{holland}.  This point is discussed in detail below, where we argue that the population size bias in DMC is likely at the root of such a discrepancy between different calculations, at least for the largest size clusters.
\hfil\break\indent
%The simulated system is enclosed in a vessel shaped as a parallelepiped of volume $\Omega$, with periodic boundary conditions in all directions; however, the size of the box is taken to be large enough (more than twice the average diameter of the cluster) to make boundary conditions irrelevant.
The quantum-mechanical many-body Hamiltonian is given by Eq. (\ref{ham}), with $\lambda=12.031$ K\AA$^2$ and  the following choice for the potential energy $V(R)$:
\begin{equation}\label{one}
V(R) = \sum_{i<j} v(r_{ij}) 
\end{equation}
Here, $v$ is the potential describing the interaction between two hydrogen molecules,
only depending on their relative distance. It should be made clear at the outset that such a simple model potential is not the most accurate choice that one could make; three-body terms are known to be quantitatively important. However, since the aim of this paper is mostly methodological, we limit ourselves to the use of a pair potential, and  select that by Silvera and Goldman \cite{sg}, for consistency with  existing calculations against which we are interested in comparing our results. We have computed ground state energies of clusters of size ranging between $N$=13 and $N$=48, using both PIGS as well as DMC.

\subsection{PIGS}\label{pigs}
Our PIGS calculations are based on a trial wave function of the Jastrow type
\begin{equation}\label{jst}
\Psi_P(R)=\prod_{i<j}\ {\rm exp} [-u(r_{ij}) ]
\end{equation}
where $r_{ij}\equiv|{\bf r}_i-{\bf r}_j|$, and
with $u(r)=\alpha/r^5$, $\alpha$ being a variational parameter whose value was set to 375 \AA$^5$ for all clusters studied here. This wave function is the same employed in previous studies based on PIGS \cite{cuervo}, albeit with a different value of the parameter $\alpha$. It is not meant to describe a finite self-bound  system, in that it only includes short-range correlations arising from the repulsive core of the intermolecular potential. A variational calculation based on such a trial wave function yields an {\it unbound} cluster, i.e., $E(N)=0$. One of the most important aspects of PIGS is precisely its ability to extract the correct physics even if the initial trial wave function is chosen less than optimally; for example, in Ref. \cite{reatto} it is shown that an accurate ground state energy estimate for solid helium can be obtained by PIGS even on setting the trial wave function equal to a constant.  This is in stark contrast to DMC, for which an appropriate choice for $\Psi_T$ often proves crucial to the accuracy and reliability of the calculation.
\\ \indent
In this work, the same approximation for $G(R,R^\prime,\tau)$ utilized in Refs. \cite{Roy,cuervo} was chosen, namely:
\begin{equation}\label{voth}
G_\circ(R,R^\prime,\tau)= G_F(R,R^\prime,\tau) 
\ {\rm exp} \biggl [-\frac{2\tau \tilde V(R)}{3}\biggr ]
\end{equation}
where $G_F(R,R^\prime,\tau)$ is the analytically known propagator for a system of non-interacting particles and
\begin{eqnarray}
\tilde V(R_{j}) = 
2V(R_{j})+\frac{\tau^2\hbar^2} {6m}\sum_{i=1}^N
(\nabla_{i}{V}(R_{j}))^2
\end{eqnarray}
if $j$ is odd, whereas $\tilde V(R_j)=V(R_j)$ is $j$ is even. It is $G(R,R^\prime,\tau)=G_\circ(R,R^\prime,\tau)+{\cal O}(\tau^5)$.
The  path sampling techniques are the same described in Ref. \cite{Roy}.

\subsection{DMC}\label{dmc}
For the DMC calculations we adopt the same trial function used in Ref. \cite{sola}, $\Psi_{D1}(R) = \prod_{i<j}{\rm exp} [-w(r_{ij})]$, where the pair pseudopotential $w(r)=\beta/r^5 +br/N$ differs from $u(r)$ of Eq. (\ref{jst}) for the need to include a linear term which prevents molecules from evaporating; the variational parameters are \cite{sola} $\beta=294$ \AA$^5$ and $b=2.79$ \AA$^{-1}$.  We also consider a much better trial function $\Psi_{D2}(R)$ with a more flexible pair pseudopotential and a three-body correlation of the standard form \cite{skl}, both optimized \cite{euler} for each cluster size.  For $N=48$, the variance of the local energy of $\Psi_{D2}$ is smaller than that of $\Psi_{D1}$ by over an order of magnitude. Significantly better trial functions can only be obtained by including four- and five-body terms \cite{nightingale}, but this route seems to be viable only for very small systems. The details of the DMC simulations are essentially those described in Ref. \cite{umrigar}, notably we utilize the standard approximation for the propagator,  supplemented with the well-known ``rejection" scheme, which has been shown to afford convergence of the numerical estimates with a significantly greater time step than would be otherwise required. 
%SM (questo undoing ora lo mostriamo esplicitamente, almeno in un caso), except that we choose to eliminate the population control bias by extrapolating to $N_W\to\infty$ rather than by undoing the bias by products of weights (see Ref. \cite{umrigar}; we come back to this point below). Also, 
We only use a slightly different translation of weights into multiplicity during the branching reconfiguration.
\hfil\break\indent
Before we discuss the results, a point must be made clear, namely that our purpose here is to carry out an unambiguous, unbiased comparison of energy estimates obtained by DMC and PIGS. 
Because we are considering a Bose system, for which the ground state wave function is positive-definite, the numerical results given by the two algorithms are expected to coincide, within statistical  uncertainties, once  extrapolations to infinite projection time for PIGS, infinite size of the population sample for DMC, and zero time step for both are carried out. Implementation details of either method,  such as the approximation adopted for the short-time propagator or the  choice of the moves in the random walk, only affect the 
efficiency of the calculations, and are of no particular concern here.  On the other hand the population  bias of DMC, which is the focus of our study, depends on the quality of the trial function (which cannot be 
arbitrarily improved in general) to such an extent 
that the extrapolation to infinite number of walkers
can be problematic or even unfeasible in practice.
\section{RESULTS}
\label{results}
\begin{figure}[t]
\includegraphics[scale=0.35,angle=0]{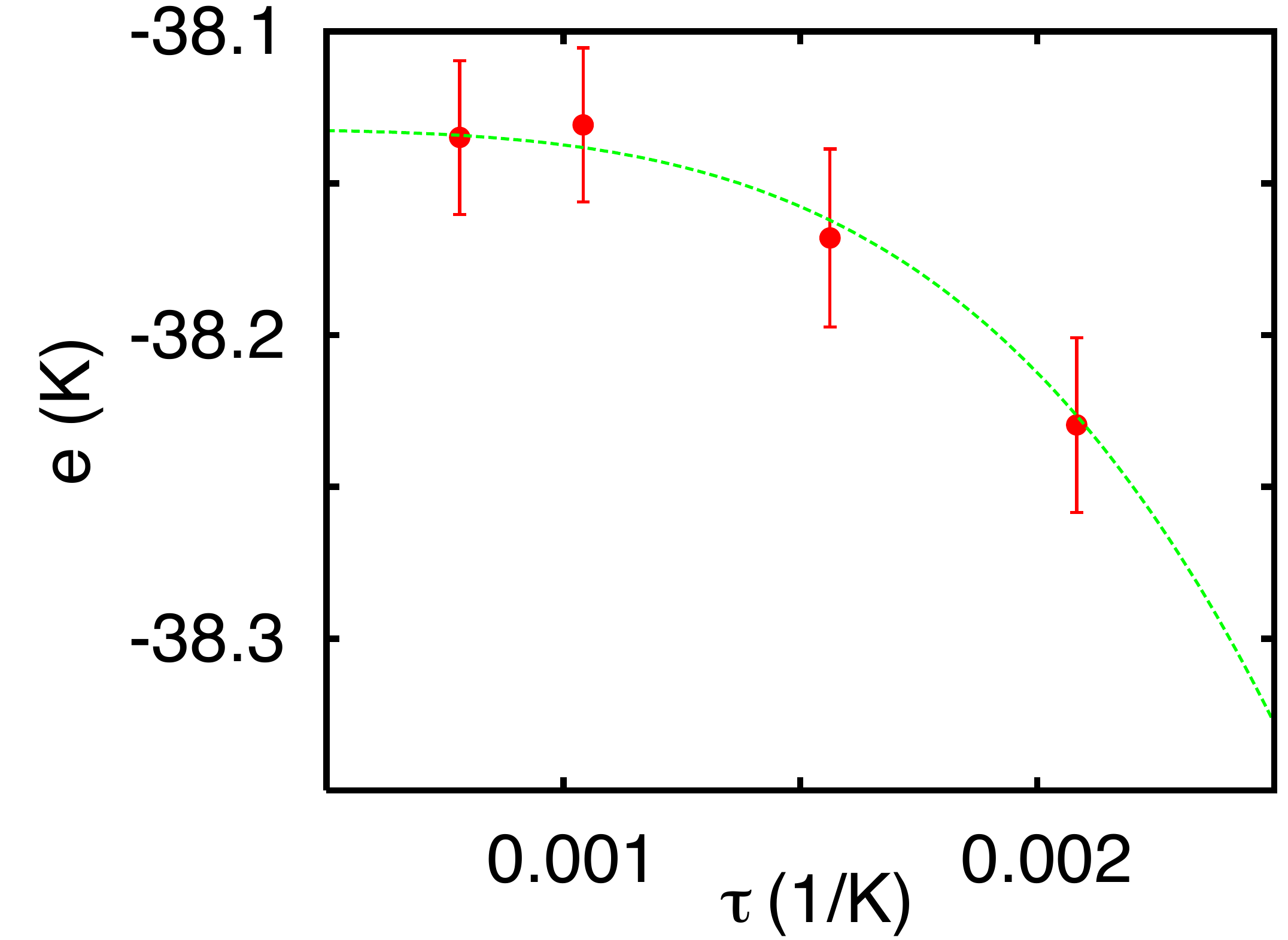}
\caption{({\it Color online}). Estimates of the ground state energy per particle for a cluster of 48
parahydrogen molecules, computed by PIGS as explained in the text, for varying values of the 
time step $\tau$ (in K$^{-1}$). The total projection time is $\beta=$1 K$^{-1}$.  
The dashed line shows a quartic fit to the data, extrapolating to
a $\tau=0$  limit of -38.14(1) K.}\label{figure1}
\end{figure}

In order to establish our main finding, we begin by illustrating results of calculations of ground state energetics for the largest cluster studied here, comprising $N$=48 parahydrogen molecules. Specifically, we compare PIGS and DMC results.\\ \indent
Figure \ref{figure1} shows estimates of the ground state energy per parahydrogen molecule $e\equiv E/N$ obtained by PIGS with a total projection time $\beta$=1 K$^{-1}$, and with different values of the time step $\tau$. A fit to the data based on the expression $e(\beta,\tau=0)=e(\beta,\tau)+c\tau^4$, justified by the use of the propagator (\ref{voth}), yields a value extrapolated to  $\tau=0$ equal to $e(\beta=1$ K$^{-1},\tau=0)=-38.14(1)$ K. 
\begin{figure}[t]
\includegraphics[scale=0.35,angle=0]{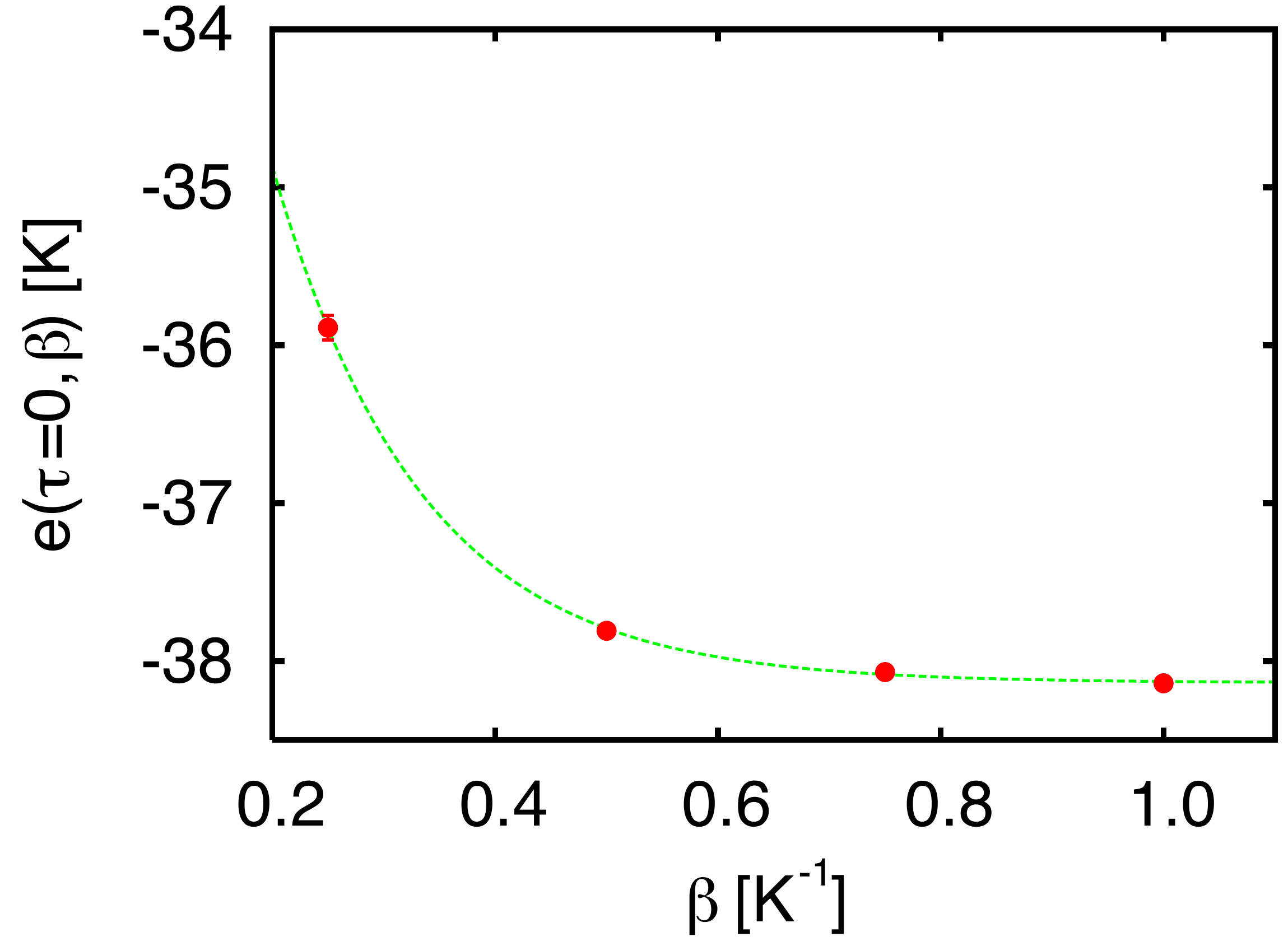}
\caption{({\it Color online}). Extrapolation to the $\beta\to\infty$ limit of estimates of the ground state energy 
per particle for a cluster of 48 parahydrogen molecules, computed by PIGS as explained in the text, for varying values of the projection time $\beta$. Estimates shown are extrapolated to
the $\tau\to 0$ limit.   Dashed line is the fitting curve described in the text.} \label{figure1b}
\end{figure}
\\ \indent
As mentioned above, an estimate for $e(\beta=\infty,\tau=0)$ can be obtained by extrapolating results obtained with different projection times \cite{curiosity}.  The result is shown in Figure \ref{figure1b}. The asymptotic value is indistinguishable, within statistical errors, from that at $\beta = 1$ K$^{-1}$. 
Our energy estimate is slightly higher than that offered in Ref. \cite{cuervo}, namely $-38.22(3)$ K,  for a projection time $\beta=0.8$ K$^{-1}$ and  with a time step $\tau$ = $1.5625\times 10^{-3}$ K$^{-1}$. For the same time step, our estimate is $-38.17 (2)$ K (see Figure \ref{figure1}), compatible with that of Ref. \cite{cuervo} if statistical uncertainties are taken into account \cite{noteourtimestep}. On the other hand, it is surprisingly almost 1 K below the most recent DMC estimate for this cluster, namely $-37.28(3)$ K,  by Sola and Boronat \cite{sola}.
\hfil\break\indent
Such a discrepancy can hardly be regarded as ``negligible", considering that the value of the chemical potential $\mu(N)$ (Eq. \ref {muc}), used to assess cluster stability, is computed by subtracting two {\it extensive} energy values, i.e., associated to whole clusters. For instance, a systematic error of the order of 0.9 K per molecule results into one  on the total energy of the $N=48$ cluster of approximately 45 K, which is  {\it very} close to the value of $\mu$ quoted in Ref. \onlinecite{sola} for this cluster.
\\ \indent
In order to shed light on this worrisome disagreement between numerical data advertised as ``exact", we have performed DMC calculations for the same cluster, as explained above.  All the results presented
so far are calculated with a time step of $2.0\times 10^{-4}$ K$^{-1}$. We find that
the time step error on the energy per particles is similar for all clusters, in the range of
sizes considered here. It does depend on the trial function, however. Our estimates
are -0.07~K for $\Psi_{D1}$ and less than 0.01~K for $\Psi_{D2}$.
\begin{figure}[h]
\includegraphics[scale=0.9,angle=0]{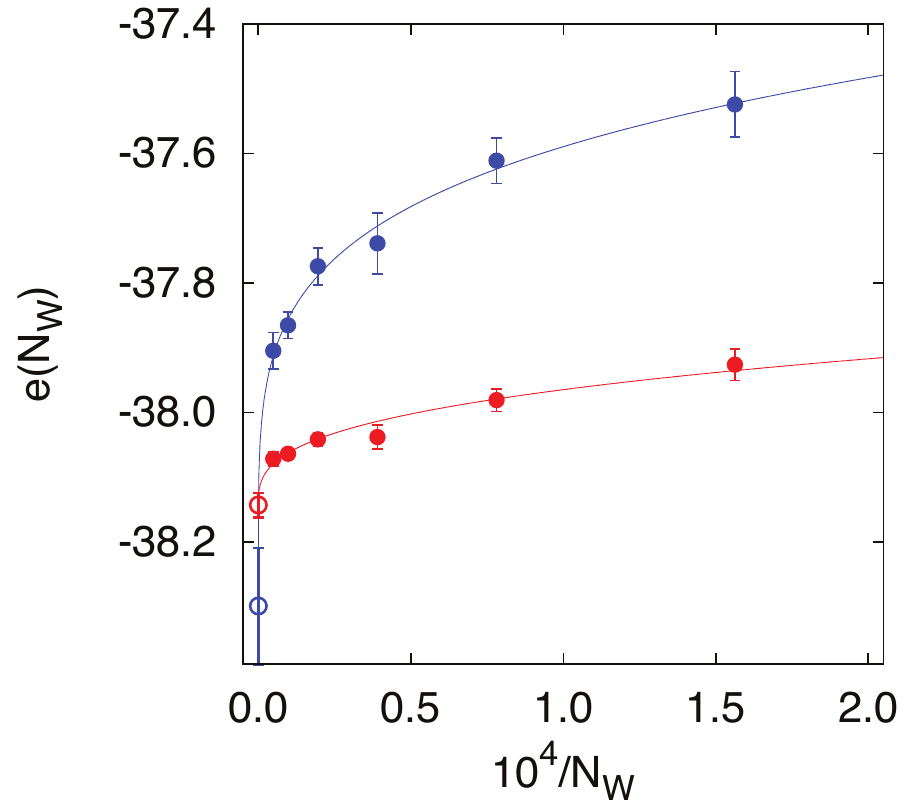}
\caption{({\it Color online}).  Ground state energy per particle for a cluster of 48
parahydrogen molecules as a function of the number of walkers $N_W$, computed 
by DMC using two different trial functions: $\Psi_{D1}$ (diamonds, blue online)
and $\Psi_{D2}$ (circles, red online). The lines are power law fits to the 
data, and the open symbols show the extrapolated values at $1/N_W=0$.
The time step utilized here is  0.0002 (in K$^{-1}$).
}\label{figure2}
\end{figure}
Figure \ref{figure2} shows the ground state energy per particle for a 
cluster of 48 parahydrogen molecules as a function of the number of walkers, 
calculated with the trial functions $\Psi_{D1}$ and $\Psi_{D2}$ of 
Sec. \ref{dmc}. If the walkers were uncorrelated, the population  
bias would vanish as $1/N_W$ \cite{umrigar}. 
This is clearly not the case: for both trial functions, we can fit data 
obtained with $N_W$ between 200 and 200,000 (not all of this range is shown 
in Figure \ref{figure2}) with the expression $e(N_W)=e(\infty)+cN_w^k$, 
and the optimal value of the exponent is 0.342 for the ``good'' trial 
function $\Psi_{D2}$, and as low as 0.202 for the ``poor'' trial function 
$\Psi_{D1}$, the reduced $\chi^2$ being smaller than 1 in both cases.
\\ \indent
There are several things to note here. First and foremost, the result 
$e=-37.278 \pm 0.028$ reported in 
Ref. \onlinecite{sola} for $N=48$, allegedly based on data ``analyzed to reduce any sistematic 
bias to the level of statistical noise'', is outside the
scale of the figure. The DMC energies of Ref. \onlinecite{sola} are systematically higher than those reported in Ref. \onlinecite{cuervo}, the difference increasing (non-monotonically) with $N$; we find it to be greatest ($\sim 0.9$ K) at $N$=48, while it is of the order of 0.2 K per molecule for $N$=30, and 0.4 K per molecule at $N$=40. \\
In Ref. \onlinecite{guardiola}, which reports DMC energy estimates (essentially identical with those of Ref. \onlinecite{sola}) for clusters of size up to $N$=40,  authors observe a ``marked" effect of population size, on performing calculations for $N_W$ ranging from 500 to 2,000, suggesting nevertheless that its overall effect on the chemical potential might be negligible, presumably due to some expected (fortunate) compensation of error.
\\ \indent
Indeed, our DMC values are similar to those of Refs. \onlinecite{sola,guardiola}, when we take $N_W\sim 1500$; the problem is that the small slope of the $e(N_W)$ curve around such a value of $N_W$ is highly deceiving, as the slope actually appears to {\it diverge}, as $1/N_W\to 0$, as clearly shown by data in Figure \ref{figure2}.  Therefore, not only is extrapolation of results which depend so dramatically 
on the number of walkers clearly problematic --  one can be easily led to believe incorrectly that convergence with respect to $N_W$ has been reached, by focusing on relatively narrow  a range of $N_W$, as in Ref. \onlinecite{guardiola}  (it does not help if discrepancies with published results by others are simply ignored).
The extrapolated value agrees, as it should, with the PIGS 
result (within two standard deviations, for $\Psi_{D2}$), 
but the amount of computer time needed to reach a given
statistical accuracy is much larger for DMC than for PIGS.
%SM questa figura non mi pare che non serva a nulla \begin{figure}[h]
% \includegraphics[scale=0.9,angle=0]{eofnw_2.pdf}
% \caption{Ground state energy per particle for clusters of 23
% (diamonds, blue online) and 48 (circles, red online) parahydrogen 
% molecules as a function of the number of walkers $N_W$, computed
% by DMC using the $\Psi_{D2}$ trial function.
% The lines are power law fits to the 
% data, and the open symbols show the extrapolated values at $1/N_W=0$.
% The time step utilized here is  0.0002 (in K$^{-1}$).
% }\label{figure3}
% \end{figure}
\\ \indent
For $N=23$ 
%SM (see Figure \ref{figure3}) 
the population bias 
is still definitely not linear in $1/N_W$, but its magnitude is much smaller than for $N=48$;  deviations from the linear behavior become hard to detect for $N=13$. We can define the number of walkers ${\bar N_W}$ needed to 
observe convergence of the energy to a precision $\epsilon$
via the relation $e({\bar N_W})-e(\infty)=\epsilon$. For 
$\epsilon=0.01$ K we find ${\bar N_W}=5000$ for $N=13$ and as much as 
${\bar N_W}=100$ millions for $N=48$ if we use $\Psi_{D2}$. A sensible 
estimate for $N=48$ using $\Psi_{D1}$ is not even possible from our
simulations, which in this case, even using up to 200,000 walkers, 
still leave a large uncertainty in the best-fit exponent of $e(N_W)$.
In terms of the comparison between the DMC \cite{sola,guardiola}
and the PIGS \cite{cuervo} results (see Table \ref{tabella}), which initially motivated this work, the dependence of 
the popolation  bias on the system size parallels and presumably 
explains the similar dependence in the observed discrepancies.
\begin{table}[h]\label{tabella}
%SM taroccato il dato a N=23 (vedi e-mail)
%   e pure l'errore a N=36, era dato a 9 centesimi, 
%   mi sa che non era cosi' grosso, ho messo 2 centesimi
\begin{tabular}{c c  c c}
$N$&DMC\cite{sola} &DMC          &PIGS        \\ \hline 
13& $-20.952(16)$  &$-20.98(1)$  &$-21.02(1)$ \\
23& $-28.111(12)$  & $-28.15(1)$ &$-28.16(1)$ \\
36 & $-33.804(19)$ &$-34.09(2)$  &$-34.13(1)$ \\
48 &$-37.278(28)$  &$-38.15(2)$  &$-38.14(1)$ \\
\end{tabular}
\caption{Ground state energy per molecule (in K) for different parahydrogen clusters, computed  by DMC (Ref.  \onlinecite{sola} and this work) and PIGS. PIGS estimates are extrapolaed to the $\tau\to 0$ limit, for a total projection time $\beta=1$ K$^{-1}$. DMC estimates obtained in this work are extrapolated to the $1/N_W\to 0$ limit as explained in the text. Statistical errors, in parentheses, are on the last digit(s).}
\end{table}

\section{Discussion}
Although the results shown above illustrate rather clearly that the bias arising from the control of the population is significant, it  could be argued that the use of  a more accurate trial wave function (e.g., $\Psi_{D2}$ instead of $\Psi_{D1}$ in the case shown in Figure \ref{figure1}),  considerably improves the convergence, and therefore it is unclear whether the problem should be ascribed to a finite population, or rather to a poor choice of $\Psi_T$.
As it turns out, although a superior trial wave function can indeed alleviate the problem of finite  population bias, this should not induce much optimism on the scalability of DMC in general. For, the behavior illustrated in
Figure \ref{figure2} is ultimately due to statistical correlation
between walkers, in turn induced by large fluctuations of the branching term
$\exp(-\tau E_L(R))$. Since $E_L$ is an extensive quantity, one can expect
--and does indeed observe \cite{nemec} -- an extremely poor asymptotic 
scaling of the efficiency of DMC with the system size. 
For molecular hydrogen, this problem compounds with a relatively
low quality of the trial function; the strength of the interparticle
potential makes it difficult to devise and use much more accurate trial wave functions 
than $\Psi_{D2}$. As a result, $N=48$ -- a very modest size for a boson 
system \cite{vermer}-- turns out to be already a demanding calculation.
\\ 
On this point, it is interesting to note that, in a previous study \cite{Roy}, a comparison of ground state energy  estimates  for bulk liquid $^4$He obtained with PIGS and DMC, found that PIGS yielded consistently lower results, and that the difference between PIGS and DMC results  increases with density. The suggestion was already made back then that the use of a finite population in DMC, comprising only a few hundred walkers in those DMC calculations, seems very likely to be the cause of the discrepancy.
\\ \indent
%SM nuova figura e relativa discussione
\begin{figure}[h]
\includegraphics[scale=0.7,angle=0]{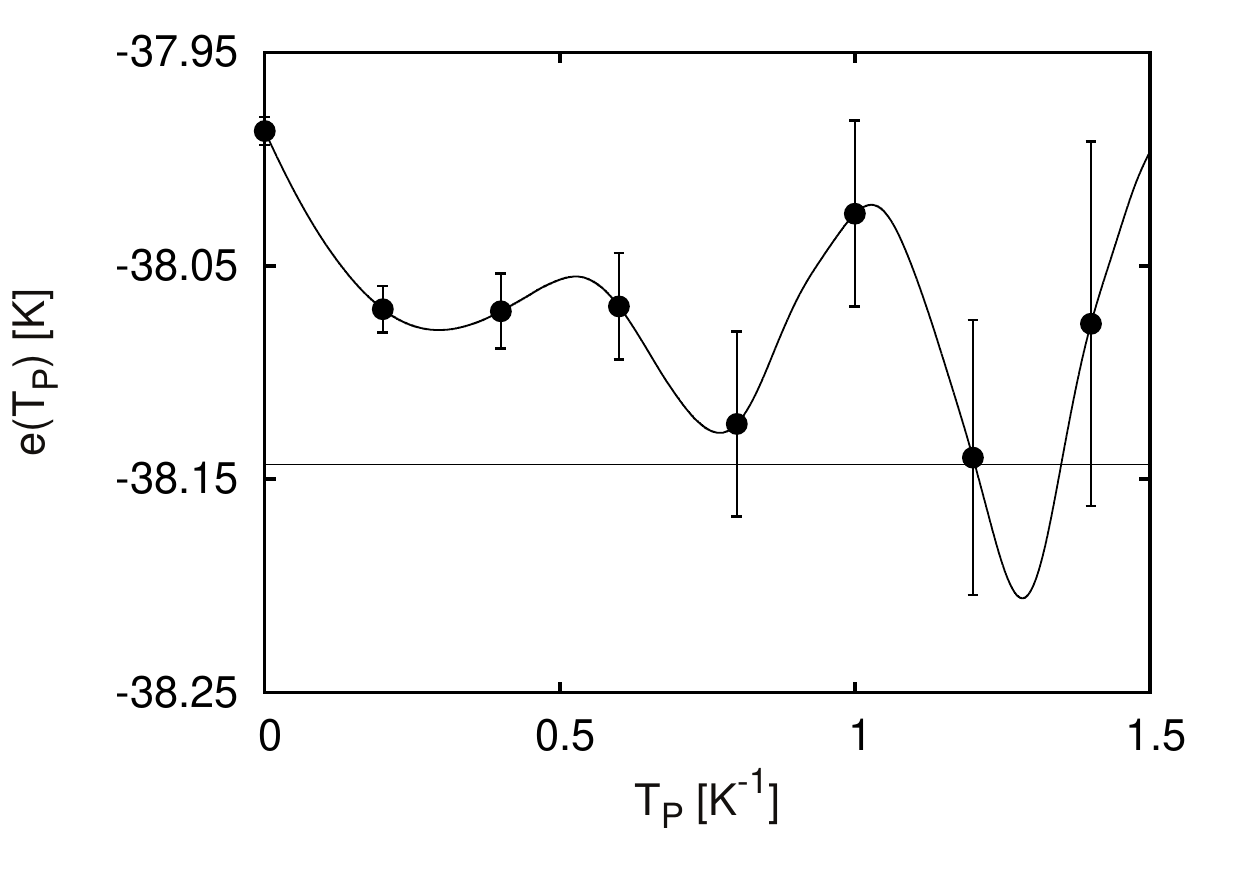}
\caption{Ground state energy per particle for a cluster of 48
parahydrogen molecules as a function of $T_P$,
computed by DMC using the $\Psi_{D2}$ trial function with $N_W=12800$.
Error bars are only shown for a few points. The horizontal line is the
extrapolated value at $1/N_W=0$ from Figure \ref{figure2}.
The time step utilized here is  0.0002 (in K$^{-1}$).
}\label{figure3}
\end{figure}
It should also be mentioned that there exists an alternative procedure,  one that in principle could remove the bias due to a finite population of walkers in DMC without requiring an extrapolation with $N_W$. One can carry out the DMC simulation with a single target value of $N_W$ and store the renormalization 
factors $f_i$
of the 
population size along the simulation \cite{umrigar}, with $i$ a time index. 
The bias would be 
eliminated by accumulating weighted averages, the weight being defined 
for each configuration
as the inverse of the product of all factors $f_i$ from the beginning of the 
simulation up to the current time. \\ While in principle this procedure completely undoes the
effect of the population control,  in practice it leads to 
unacceptably large variance. Thus, one keeps in the weighted averages only the product of 
the last $K_P$ factors $f_i$, and seeks convergence of the results by 
increasing the ``correction time'' $T_P=K_P\tau$. 
However, one is bound to face severe efficiency problems whenever 
the population size bias is strong. Figure \ref{figure3} illustrates an
attempt at correcting the population size bias for the ground state
energy of a cluster of 48 parahydrogen molecules calculated with 12800 walkers. 
From the biased value at $T_P=0$ the energy 
is expected to converge for large times to the exact value (the extrapolation
of Figure \ref{figure2}, shown in Figure \ref{figure3} by the horizontal line).
However no convincing evidence of convergence can be detected before 
the statistical error grows as large as the bias itself, despite this 
simulation being 8 times longer than that performed for the single point 
at $N_W=12800$ of Figure \ref{figure2}.
\\
\indent
In conclusion, we have presented numerical evidence to the effect that the bias arising from a finite  population size in DMC calculations is the most likely cause of discrepancies reported in the literature between ground state energy estimates for Bose systems obtained with DMC and Metropolis-based methods such as PIGS. Although a complete removal of the bias  (whose magnitude appears to have been generally underestimated, or in any case not fully appreciated) is possible in principle, the computational resources required grow significantly with system size. In fact, although the system sizes for which we are presenting data in this work are too small to make that conclusion, they are strongly suggestive of exponential scaling.  Obviously, although we have illustrated quantitatively this conclusion on a Bose system, it applies equally to fermions, there being nothing in the argument expounded here that depends on quantum statistics. If anything, there are reasons to expect that the use of the popular fixed-node approximation to circumvent the sign problem may conceivably worsen the problem of fluctuating local energy, which is at the root of the population bias. 
Thus, while the choice between the two methods has been so far largely regarded as one of ``personal taste",  path integral methods, requiring no walker population, may prove a better choice for systems of large size, 
%SM
how large depending
on the quality of the trial function.
\\
\indent
Finite temperature methods such as Path Integral Monte Carlo, which do not require a population of walkers, also do not suffer from the kind of bias discussed in this work, that affects instead any population-based procedure such as GFMC (including for lattice Hamiltonians) and DMC. Thus, although one may naively think that ground state methods would necessarily be better suited for $T$=0 calculations, PIMC may in fact  also prove a better option than DMC in some cases, given the significance of the population size bias. It is worth mentioning that for the specific physical system discussed her, PIMC yields estimates in the $T\to 0$ limit consistent with those furnished by PIGS \cite{holland,mezzacapo, mezzacapo1}.
\section*{Acknowledgments}
This work was supported in part by CASPUR under HPC grant 2012. Useful discussions with Fabio Mezzacapo are gratefully acknowledged.

\bibliography{paper}

\begin{thebibliography}{10}
\bibitem{notesign}
Population bias affects DMC regardless of quantum statistics, and there are no obvious reasons to expect it to be worse for either Fermi or Bose systems. Of course, QMC simulations of Fermi systems are also affected by the ``sign problem", but this is a separate issue, unrelated to what we discuss in this paper. We therefore restrict our discussion to Bose statistics for convenience and simplicity.
\bibitem{notey}
Strictly speaking, $\Psi_T$ is required to be non-orthogonal to the true ground state wave function. For a Bose system 
(such as condensed $^4$He) this is not a problem, as  the ground state wave function can always be chosen real and 
positive, and therefore any positive-definite function $\Psi_T$ satisfies the non-orthogonality requirement. That $\Psi_T$ be non-negative is of course also 
crucial in order for (\ref{pippo}) to be treated as a probability.
\bibitem{noteourtimestep}
The values of the energy per molecule published in Ref. \cite{cuervo} appear to be affected by a systematic downward shift worth between 0.05 and 0.1 K, due to time step error. We have established that, within PIGS, the largest time step for which estimates are indistinguishable from those extrapolated to the $\tau\to 0$ limit, within our quoted statistical uncertainties, is approximately $1.0\times 10^{-3}$ K$^{-1}$.
\end{thebibliography}
\end{document}